# Enhancing Public Service Delivery through Organisational Modelling

**Anna Marie Fortuito, Chandana Withana**
School of Computing and Mathematics,
Charles Sturt University, Sydney, NSW 2010, Australia
Email: amlfortuito@gmail.com, cwithana@studygroup.com

**Moshiur Bhuiyan, Farzana Haque, Luba Shabnam**
Service Consulting,
Enterprise Cloud Systems Pty Limited
Sydney, NSW 2560, Australia
Email: {moshiurb, farzana, luba}@ecloudsys.com

**Aneesh Krishna**
Department of Computing
Curtin University, Perth, WA 6102, Australia
Email: a.krishna@curtin.edu.au

## Abstract

Government organisations utilise a Citizen's Charter in order to furnish the public with comprehensive details of their service offerings that defines how their overall organisational goals are achieved. Studies on the implementation of Citizen's Charter in developing countries indicated that important social factors were not considered. Reviewed literatures indicated that social factors could accelerate service delivery with the use of goal-oriented modeling (such as *i** organisational modeling) as it represent conditions expected from social actors and their social dependencies. By modeling role dependencies among actors carrying out services in a business unit and applying vulnerability and criticality metrics, problem service areas were identified using the Citizen's Charter as a source of these methodologies. Improvements in these problem service areas were based on vulnerability and criticality levels of their corresponding actors. Recommendations to address these service areas include monitoring of key performance indicators of actors and task delegation when necessary.

## 1 Introduction

Government organisations share a common goal to improve the living standards of their citizens through public service offerings. Like the private sector, government organisations eagerly implement systems for more suitable operational intensities. The viability of systems is based on its coordination with an intended environment.

e-Government has been used in recent years to intensify the interaction of governments with other government agencies, business organisations, its employees, and its citizens as it promoted stakeholder contribution to development and expanded governance process through the use of the Internet (Hai 2007). As a result, official websites of government organisations has emerged, primarily publishing pertinent information that the public requires. Further, some of these websites are equipped with information systems that serve as a convenient avenue for citizens to avail some services. Government organisations that have adapted policies of public transparency and disclosure also feature procedural steps on how to avail services in the form of a Citizen's Charter, in conjunction with their manual and information systems.

A Citizen's Charter is a publicised document that defines the type, quality, and magnitude of service that the citizenry can expect from his government (Mang'era and Bichanga 2013). It describes services offered consisting of employees with their responsibility to carry out a specific service, clients with actions expected from them to do to avail the service, requirements that the client needs to provide to employees, procedures of the service, and fees to be paid if applicable. Any issue that these factors will encounter delays the expected duration of a service, weakens the quality of the service, and eventually creates an overall adverse impression to the public. As the Citizen's Charter summarises the overall goal of a government organisation as well as describe how specific business unit's service goals are attained, it can serve as basis in understanding and capturing how people, with varying intentionality, are dependent on business processes.

Business process details such as activity sequence and decisions are predominantly used for the development of information systems (Loucopoulus and Kavakli 1995). As much as it achieves the performance and simulation of information systems, there is an absence of the representation of social and intentional components that can help out in process redesign and enhancement (Yu 1995).

Organisational modelling can be used to model intentionality and interdependency of actors reliant on business processes. Association of organisational models with business processes during the early phase of systems development can address social issues that affect the implementation of systems. In addition, organisational modelling can assist in system's planning that can be incorporated in the government organisation's administration and strategy (Frank 2015).

This paper aims to establish how organisational modelling can enhance public service delivery. Using a Citizen's Charter as basis for modelling, this paper contributes on leveraging requirements analysis by providing a methodology to identify probable limitations caused by actors and ultimately recommends means to tackle it.

As developing countries require more improvement in terms of public service delivery, this paper targets to study Marikina City, a local government unit in the Philippines. The rest of this paper is organised as follows. Section 2 presents the background of this study. Section 3 describes the *i** organisational modelling framework. Section 4 discusses the methodology undertaken. Finally, Section 5 presents conclusion and future works.

# 2 Background

An organisation's capability depends on how the mixture of its resources such as people, business processes, assets, and technology solutions support its predetermined goals (Grant 1991; Kangas 2003). In realising and sustaining these goals, organisations must adapt to its dynamic environment through a continuous improvement of its resources. Modelling organisational goals is a fundamental approach in understanding how organisation's resources depend on each other to achieve goals and to identify, which among its resources need improvement.

Various modelling languages like Data Flow Diagram (DFD) and Unified Modelling Language (UML), have been used in previous years to model business processes. These modelling languages respectively use the "what" method to capture business processes and the "how" method that decomposed a process into more specific details (Satzinger et al. 2012). Though these modelling languages illustrated how goals can be achieved, they do not exhibit alternatives to accomplish a goal. Feather (1987) introduced the concept of a 'why' method that presented the manner actors with varying intentions can influence a process in attaining goals. Other authors (Yu et al. 2010) later acknowledged the greater importance of actors rather than processes in goal achievement. Actors or stakeholders have autonomous intentions, and coexist with other actors in an environment. When the intentions of actors to the environment are identified, various alternatives can be determined.

An actor is a predominant notation in goal-oriented modelling. Goal oriented modelling represents conditions expected from an actor with its dependency relationships to other actors. It deals with interactions among actors and their influence to process goals. Goal-oriented modelling can be accomplished with the use of several notations like *i**, KAOS (Knowledge Acquisition in Automated Specification), and Tropos. According to the evaluation of researchers (Matulevičius et. al 2006), *i** is more appropriate in capturing initial requirements compared to other goal-oriented models. In gathering requirements, understanding procedures to carry out a service is not enough and that reasoning of stakeholder's goals should also be taken into account (Yu 1997). *i** can demonstrate an organisation with its stakeholders, the influence of each stakeholder, and how services can be achieved.

In the achievement of organisational goals, it is imperative that organisations can identify important service areas to demonstrate that they are on top of things. Bhuiyan (2012) proposed an intuitive approach in analysing the strategic dependencies among actors in order to measure and identify each actor's vulnerability and criticality. According to this author, outgoing dependencies is indicative of task delegation. An actor becomes vulnerable if tasks, goals, and resources are delegated to another actor. Hence, if these factors are not satisfied, the actor becomes vulnerable. On the other hand, actor's criticality measurement is based on incoming dependencies that are deemed responsibilities of the actor. If an actor fails to satisfy incoming dependencies, the actor becomes critical. This paper furnishes a new application of goal-oriented modelling with the use of *i**. This framework is applicable for government organisations that have Citizen's Charter.

# 3 *i**Organisational Modelling Framework

*i** is an organisational modelling framework that can show organisational context and can answer questions like how an actor depend on other actors, what are existing goals, and what alternatives are there to be considered (Yu et al. 2010). Knowing the modelling elements and applying them in the model is imperative to be able to construct a framework that is in actual representation of the organisation (Vazquez et al. 2013). The concept of *i** is centred on intentional actors. The actor represents intentional characteristics along with dependencies like goals to accomplish, tasks to do, resources to provide, and soft goals (objectives and preferences) to satisfice. Actors can be an organisation, tangible/intangible entities, and agents. A goal is a condition, desire or interest. Soft goals are desires or interests that can be vague in the satisfaction criteria. Tasks are distinct actions or activities. A resource is a touchable object or informational entity. Figure 1 illustrates *i** organisational modelling notation's basic structural elements.

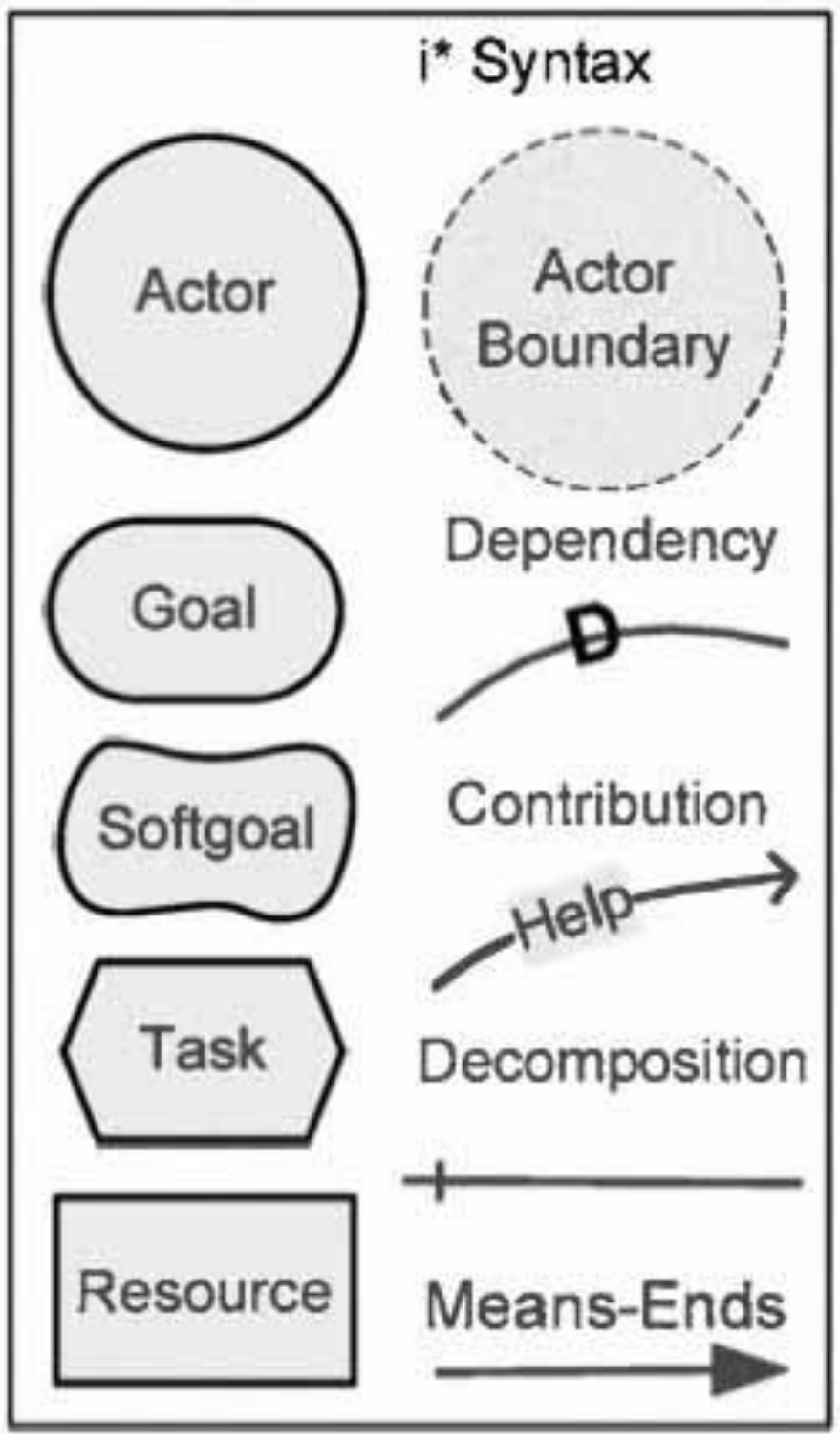

*Figure 1: i* Syntax*

*i** framework comprise of two main components namely Strategic Dependency (SD) and Strategic Rationale (SR) that illustrates the world based on perceptions. These components contain nodes and links. A node signifies an actor and a link between actors implies that one actor depends on the other for the former to achieve a goal. The depending actor is categorised as the 'depender' whereas the actor depended upon is the 'dependee'. The object that establishes the dependency relationship is termed as 'dependum'.

An SD model is a representation of intentional points that symbolises actors as well as dependencies that denote relationships (Ayala et al. 2005). In constructing the SD, initial analysis of the domain is suggested to make the goals clearer, and then create goal dependencies among actors (Franch et al. 2007). With the many challenges in domain analysis, it is imperative to identify strategic, business, and information goals in parallel with levels of abstractions as these are related to decisional goals (Franch et al. 2011). An SD model example is illustrated in Figure 2 for a birth registration service. Outgoing arrows from an actor denotes that the actor is the depender. Incoming arrows to an actor denotes that the actor is the dependee. The element between actors is the dependum. To interpret the model, the Registration Officer is dependent on the Customer to provide a resource (birth registration requirements) and do a task (present registration fee payment). Registration Officer is the depender and the Customer is the dependee. The goal in this example is the processed on time birth registration where the dependency relationship between the actors was reversed. Registration Officer became the dependee and the Customer became the depender. The Customer also has dependency relationships with the Cashier.

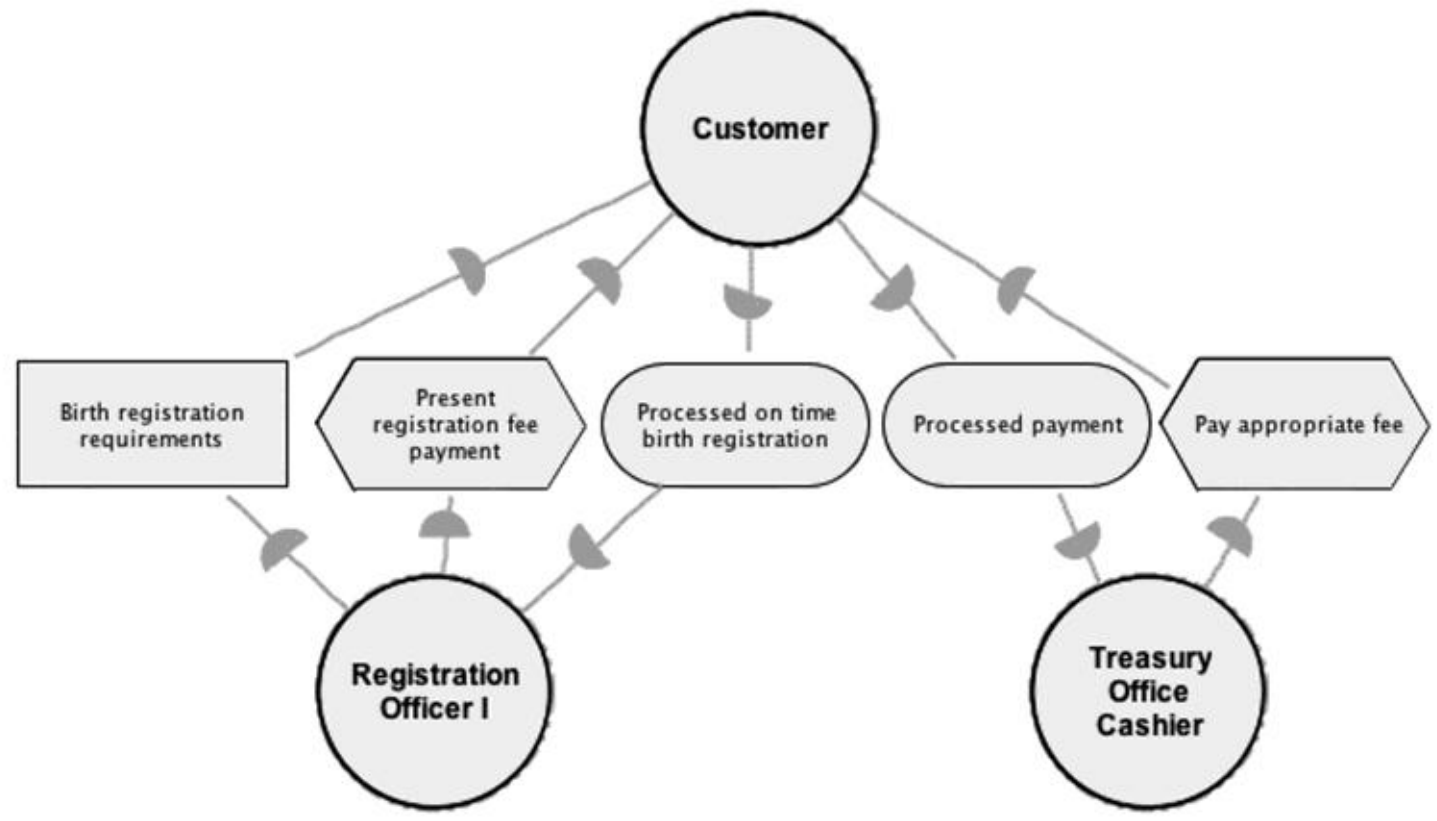

*Figure 2. SD Model for Birth Registration Service*

An SR model magnifies the SD model by disintegrating its elements with task decomposition and means-end link. Task decomposition furnishes details on a task and its hierarchical itemisation to subtasks. Means-end links connect goals to the tasks and resources required to achieve them. Task decomposition and means-end links are to be performed by each actor and are enclosed in the actor's boundary. Figure 3 illustrates an SR model of the same service area. The Registration Officer has an internal task to process birth registration. This task is performed into subtasks (verify requirements and advise missing requirements), and resource (requirement checklist).

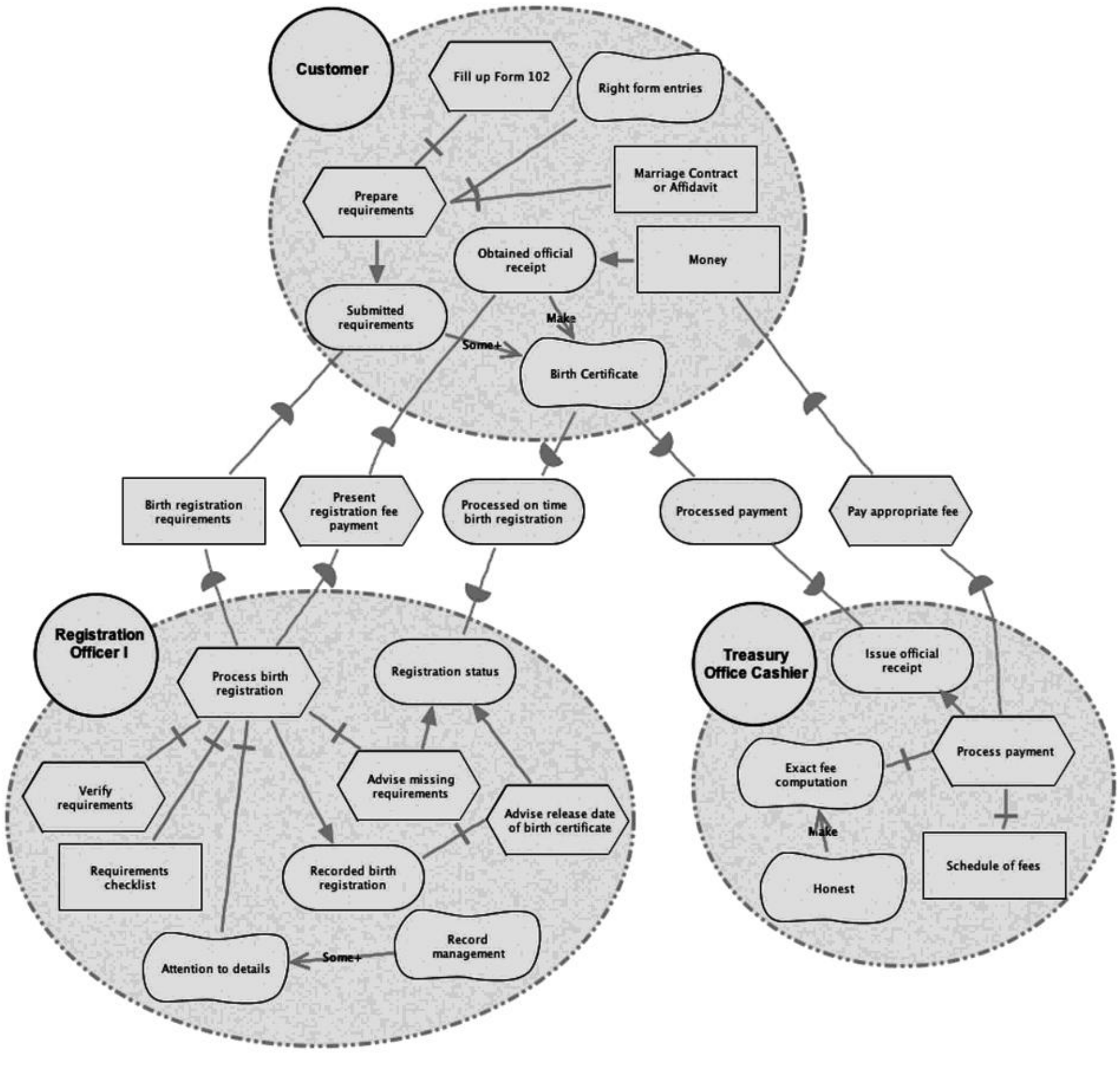

*Figure 3. SR Model for Birth Registration Service*

# 4 Methodology

The methodology used in this paper is composed of three main steps. The first step is to utilise a Citizen's Charter to perform SD modelling. The Citizen's Charter of the City Civil Registrar Office of Marikina City is used for this purpose. The second step is to measure actors' vulnerability and criticality based on the SD model. Finally, step 3 is the analysis of results and propose enhancement.

## 4.1 *i**Modelling in the Context of Citizen's Charter

A Citizen's Charter comprises the description of responsibilities of all the business units in a government organisation. Per business unit, all its service areas are presented with requirements and procedures on how to avail the service. Per procedure, there is a clear description of what a service recipient and service provider needs to perform and its expected duration.

To start off the modelling, various actors must be identified. An actor is the dominant modelling construct, as they are considered as being intentional, having strategic interests and goals. Actors are found in the "person in charge" of a particular procedure. Per actor, identify goals that represent the attainment of a certain service. Per goal, identify resources needed and tasks to be performed by the actor. Resources represent requirements needed for a specific service. Tasks are action undertaken by the actor. Per resource, task, and goal, identify the other actor involved. Actors have several dependencies to each other so as to work jointly to attain their goals. Once all of these are identified, represent them using *i** notations. This will highlight the analysis of actors' cooperation to achieve domain goals through dependencies (Esfahani et. al 2010).

The Citizen's Charter of the City Civil Registry Office of Marikina City embodies a comprehensive description of the services they offer namely issuance of certified copy of civil registry documents (birth, marriage, and death certificates), on-time and late registrations, issuance of marriage license, registration of legal instruments, registration of court decrees, registration of foundling, and endorsement of documents to the National Statistics Office. These services are delivered by their staff to meet the ultimate goal of the office that is "to record and safe keep vital events and other documents where acts, events, legal instruments and court decrees affect the civil status of the people" (Marikina City 2012).

There are 9 front line employees (Registration Clerk/Verifier, 4 Local Civil Registry Clerks, Assistant Registration Officer, Registration Officer I, Registration Officer II, and Registration Officer III) catering to the needs of service recipients. Each of these staff performs multiple roles in carrying out one or several services. The City Civil Registry Office interacts with 2 other internal offices of Marikina City (Treasury Office and Health Office) and 4 external offices (Post Office, National Statistics Office, Local Courier Personnel, and other local civil registry offices). All service recipient groups are represented as Customer seeing that they are a single identity availing various services. Overall, there were 16 actors found that interact among each other.

Figure 4 illustrates the SD model of the City Civil Registry Office incorporating all their services. In the SD model, the Customer actor represents all service recipient groups hence the Customer actor has dependency relationships in almost all the other actors. This model features how citizen-centric the services are. Dependency elements of each actor are based on the roles they play for a particular service.

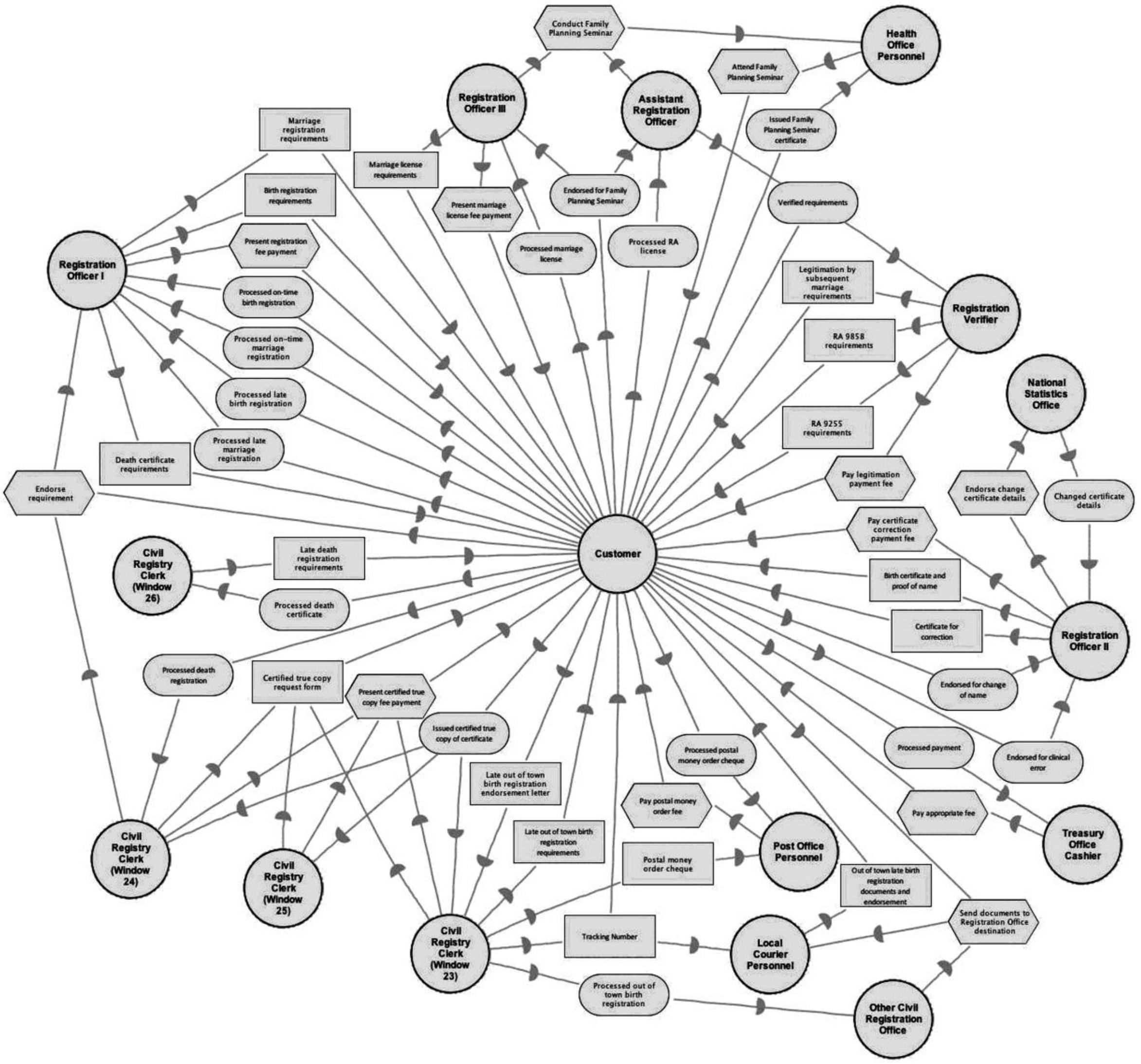

*Figure 4. SD Model of the City Civil Registry Office Based on the Citizen's Charter*

## 4.2 Measuring Actors' Vulnerability and Criticality

It is imperative for organisations to identify important service areas objectively. The developed SD model illustrated a network of dependency relationships among actors (Yu 1995) that can identify which actors are vulnerable and critical. This can be done by coming up with vulnerability and criticality measurements of actors. These measurements increase if an actor is associated with more vulnerable and critical actors (Islam et al. 2009; Shabnam et al. 2014). A service area that the actor is responsible with is carried if the actor has high measurements of vulnerability and criticality.

In this paper, the formula used was all based on the proposed metrics of Bhuiyan (2012). Vulnerability and criticality measurements will only cover actors within the business unit. The Customer, Post Office Personnel, Local Courier Personnel, National Statistics Office, and Other Civil Registry Office will not be included in the vulnerability and criticality measurements. The Health Office Personnel and Treasury Office Cashier belonging to other business units will not be included as well. These excluded actors are important to be represented in the SD model as they contribute to the actual vulnerability and criticality measurements of actors belonging to the studied business unit.

### 4.2.1 Vulnerability

Vulnerability is the exposure to threats of anything that has value to the organisation (Rosado et. al 2013). In this context, the vulnerability of actors is subject to outgoing dependencies that denote delegation of the different dependency elements of the depender to the dependee. A depender's rationality is to have resources available, tasks performed, and goals achieved by the dependee. If the depender is successful in obtaining dependum, it is less likely that he becomes vulnerable. The number of outgoing dependency is a major factor in measuring the vulnerability of an actor. When there are more dependees, the lesser are the chances of vulnerability. This denotes that if an actor has many outgoing dependencies but also has many dependees, the less possibility the depender becomes vulnerable. The more outgoing dependencies an actor have with lesser dependees, the greater the chance of vulnerability. This paper used the formula below to measure actor vulnerability.

VM= number of outgoing dependencies / number of dependee actors

The table below illustrates the vulnerability measurement of actors based on the SD model developed.

| Actors | # of Outgoing Dependencies | # of Dependee Actors | VM |
|---|---|---|---|
| Registration Officer I | 4 | 1 | 4.0 |
| Registration Officer II | 4 | 2 | 2.0 |
| Registration Officer III | 3 | 2 | 1.5 |
| Assistant Registration Officer | 2 | 2 | 1.0 |
| Registration Verifier | 4 | 1 | 4.0 |
| Registration Clerk (Window 23) | 6 | 3 | 2.0 |
| Registration Clerk (Window 24) | 3 | 2 | 1.5 |
| Registration Clerk (Window 25) | 2 | 1 | 2.0 |
| Registration Clerk (Window 26) | 1 | 1 | 1.0 |

*Table 1. Actors' Vulnerability Measurement*

### 4.2.2 Criticality

Criticality is a measure to identify who among the actors has the most number of constraints (Morandini et. al 2009). In contrast with vulnerability, the criticality of an actor is based on incoming dependencies that denote responsibilities allocated to him. In measuring criticality, the actor concerned is the dependee. If the dependee succeeds in satisfying incoming dependencies, the less likely he becomes critical. This paper used the formula below to measure actor criticality. The formula denotes that the more number of incoming dependencies and depender an actor has, the greater the criticality.

CM = number of incoming dependencies * number of depender actors

The table below illustrates the criticality measurement of actors, also based on the SD model developed.

| Actors | # of Incoming Dependencies | # of Depender Actors | CM |
|---|---|---|---|
| Registration Officer I | 5 | 2 | 10 |
| Registration Officer II | 3 | 2 | 6 |
| Registration Officer III | 2 | 1 | 2 |
| Assistant Registration Officer | 2 | 1 | 2 |
| Registration Verifier | 1 | 1 | 1 |
| Registration Clerk (Window 23) | 2 | 1 | 2 |
| Registration Clerk (Window 24) | 2 | 1 | 2 |
| Registration Clerk (Window 25) | 1 | 1 | 1 |
| Registration Clerk (Window 26) | 1 | 1 | 1 |

*Table 2. Actors' Criticality Measurement*

### 4.3 Result Analysis and Enhancement Proposal

The SD model developed based on the Citizen's Charter of the City Civil Registry Office showed various actor dependencies in carrying out all the goals or services of the business unit. It illustrated how these goals were centred towards the Customer actor that represented various service recipients. It also illustrated the distribution of services, as represented by goal dependency elements in the SD model. At first glance, it is apparent that some actors (like the Registration Office I, Registration Officer II, and Civil Registry Clerk in Window 23) have too many dependency elements compared to the rest of the actors. However, to state that there is improper distribution of services based on the number of dependency elements per actor cannot be justified by merely looking at the model. The vulnerability and criticality measurements were used to accurately identify who among the actors in the business unit needs attention.

Based on the results in Table 1, it was identified that the Registration Officer I and Registration Verifier actors are the most vulnerable. These two actors are the most vulnerable because they have 4 outgoing dependencies with just 1 dependee compared to other actors who have more outgoing dependencies but with more dependees. This indicates that these two actors are vulnerable because they are only depending on a singular dependee to obtain their 4 dependency elements, which classifies them to be more vulnerable compared to the others.

In the criticality measurement as shown in Table 2, the Registration Officer I again was identified the most critical. The Registration Officer I actor has 5 incoming dependencies and 2 dependers compared to other actors who have lesser incoming dependencies and dependers. This implies that since more dependers are relying on the actor to achieve the dependency elements, this makes the actor more critical in comparison with the other actors.

The developed SD model offered a source to explore implications of alternative organisational changes in terms of management and structure. In tackling organisational change, it is imperative to consider who among actors does what, how, and why (Rolland and Grosz 1994). The SD model illustrated patterns of relationships that can temper vulnerability and criticality. Having identified the most vulnerable and critical actors, it is comprehensible which service areas needs improvement. Improvements in this context cover how service areas can be strengthened through performance monitoring and delegation of responsibilities when necessary in order to reduce vulnerability and criticality measurements.

Every actor contributes to the overall attainment of organisational goals. An actor's vulnerability is measured based on incoming dependency elements and dependees. This implies that an actor is dependent on these factors to achieve his designated goals. In order to discern if an actor's vulnerable state is affecting the organisation, there must be a tangible measure to track his performance against his designated goals (Dalziell and McManus 2004). Hence, it is suggested that monitoring of tasks be undertaken for vulnerable actors. When performance indicators go below benchmarks, it pinpoints that the vulnerable state of actor is already compromised and further action is needed to refine the delivery of the service.

The increase of the criticality state of an actor is determined on the achievement of his outgoing dependencies. Failure to attain these dependencies will eventually affect the performance of the depender. Hence, it is imperative that the performance of the depender will not be disrupted. This can be addressed through allocation of the goal to another knowledgeable actor (Pfeffer and Salancik 2003). This denotes that if there is an increase in the criticality state of an actor due to numerous incoming dependencies and dependers, clustered dependencies must be delegated to another actor who is knowledgeable enough to achieve the goal of the service area.

Approaches in addressing vulnerability and criticality can be combined when there is urgency to restructure the management of goals in the business unit. The Registration Officer I actor is both vulnerable and critical. It is important to monitor first the performance of the Registration Officer I in obtaining his goals. Should the performance indicator of Registration Officer I falls short of the target, it is then suggested to delegate clustered dependency elements to another actor who is knowledgeable of the goal. In this case, it is the Registration Clerk in Window 26 actor. This actor's goal focuses in the late death registration service, which is similar to one of the service responsibilities of the Registration Officer I actor. Delegating these specific clustered dependencies to Registration Clerk in Window 26 actor will balance the criticality and vulnerability measurements. The same principles apply to the Registration Clerk actor where delegation of his outgoing dependencies can be directed to the Assistant Registration Officer actor. However, delegation is not applicable when it will make other actors the most vulnerable or critical. Such is the situation of the Registration Officer II actor in case delegation will be made to the Assistant Registration Officer actor. For cases like this, it is then necessary to supplement the number of manpower or actors in the delivery of the service.

Table 3 below summarises changes to the measurement of actors who are most vulnerable and critical. One outgoing and incoming dependency of the Registration Officer I actor was delegated to the Registration Clerk in Window 26 actor, as well as 1 depender. Though Registration Officer I is still the most vulnerable actor, there is already no large gap among all vulnerability measurements. The same was applied to the Registration Verifier actor to the Assistant Registration Officer, but only affecting the vulnerability measurement. Two outgoing dependencies were allocated to the Assistant Registration Officer actor without adding the number of dependee actors since the Customer actor is the already an existing dependee.

| Actors | # Outgoing Dependencies | # of Dependee Actors | VM | # of Incoming Dependencies | # of Depender Actors | CM |
|---|---|---|---|---|---|---|
| Registration Officer I | 3 | 1 | 3 | 4 | 1 | 4 |
| Registration Officer II | 4 | 2 | 2 | 3 | 2 | 6 |
| Registration Officer III | 3 | 2 | 1.5 | 2 | 1 | 2 |
| Assistant Registration Officer | 4 | 2 | 2 | 2 | 1 | 2 |
| Registration Verifier | 2 | 1 | 2 | 1 | 1 | 1 |
| Registration Clerk (Window 23) | 6 | 3 | 2 | 2 | 1 | 2 |
| Registration Clerk (Window 24) | 3 | 2 | 1.5 | 1 | 1 | 1 |
| Registration Clerk (Window 25) | 2 | 1 | 2 | 1 | 1 | 1 |
| Registration Clerk (Window 26) | 2 | 1 | 2 | 2 | 2 | 4 |

*Table 3. Proposed Vulnerability and Criticality Measurements*

The recommended changes in vulnerability and criticality measurements will position the Registration Officer II as the most critical factor. Should delegation be made to the Assistant Regional Officer actor, being the most qualified for the delegation, it will make the Assistant Regional Officer actor the most vulnerable. Since the Assistant Regional Officer actor is already in the safe zone of both measurements, the delegation was not recommended. Hence the criticality position of Registration Officer II actor is in status quo until such time that management decides to increase its number of actors and delegate dependencies of the Registration Officer II actor to a new actor.

With movement indication of incoming and outgoing dependencies including a depender in the metrics to balance vulnerability and criticality measurements, improvement of service areas were involved as well. This denotes that having addressed the issues of vulnerability and criticality, problem service areas were also solved. This is based on the concept that actors carry with them the responsibilities in carrying out a goal that represents a service. The following problem service areas has been identified:

- Death registration (as per service offered by Registration Officer I actor);
- RA 9858 (as per service offered by Registration Verifier actor); and
- RA 9255 (as per service offered by Registration Verifier actor).

Having identified problem services areas based on the vulnerability and criticality metrics, the said problem areas need to be delegated to other actors. This approach can assure that these problem service areas has more probabilities of delivery if responsibilities are delegated to actors that are less vulnerable, critical, and has the knowledge in carrying out the service. Based on the suggested vulnerability and criticality metrics, Figure 5 is the proposed SD model of the City Civil Registry Office. The said model illustrates adjustments in terms of delegations made in the vulnerability and criticality measurements as highlighted by actors and their corresponding dependency elements.

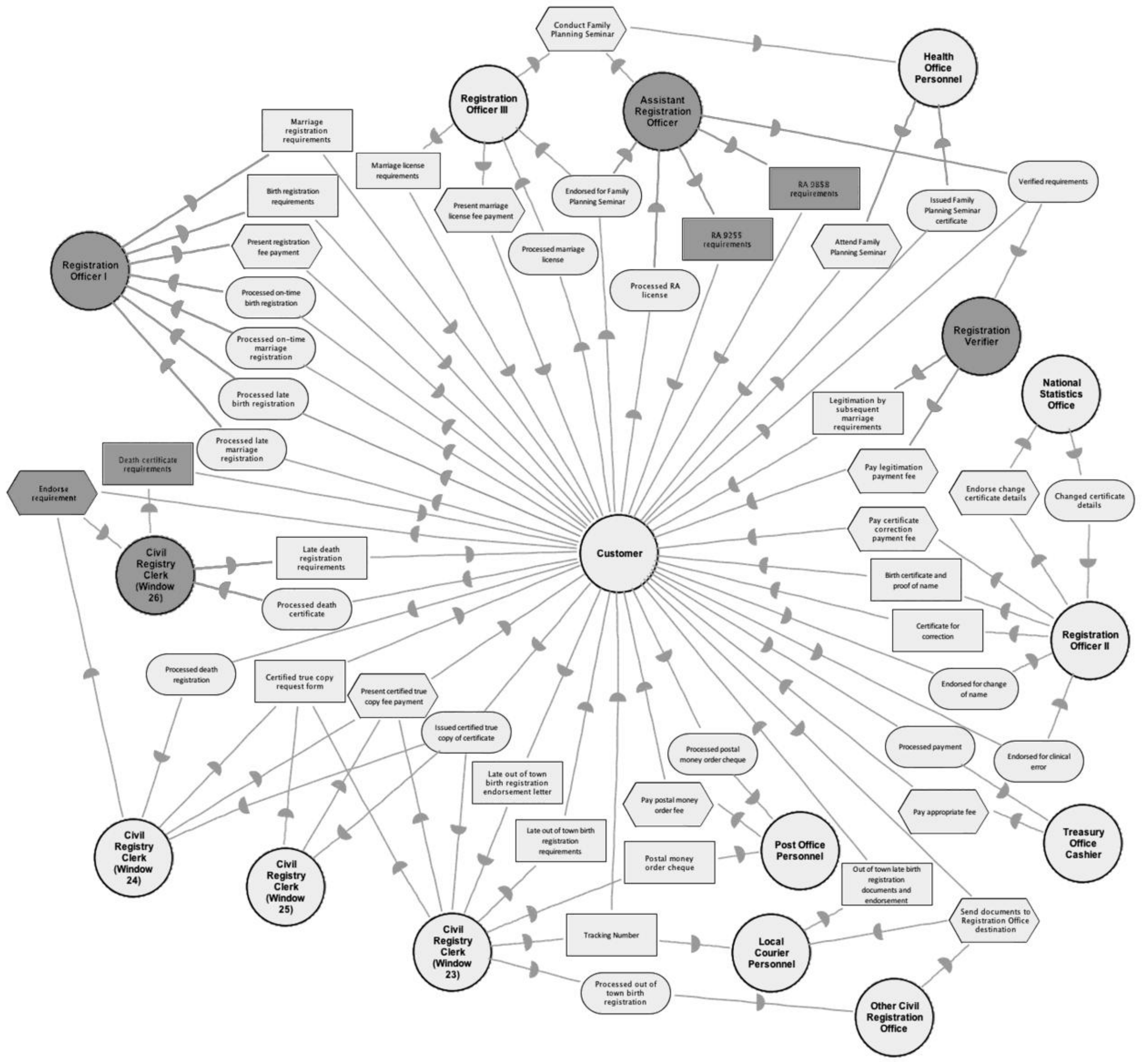

*Figure 5. Proposed SD Model of the City Civil Registry Office*

# 5 Conclusion and Future Work

This paper illustrated a methodology to improve service delivery through the use of organisational modelling. The improvement suggestions were based on analysing a Citizen's Charter and developing a based model that represents role dependencies within the organisation in service delivery context. The strategic dependency and intent was taken into consideration while measuring vulnerability and criticality of service areas. Model validation of the artefacts within this work was not in scope, as current literature was deemed adequate for this purpose.

Operation process management and modelling improvement is not in scope of this work. Even though Citizen' Charter defines the step by step procedures in carrying out services to attain business goals, further requirements gathering, such as role qualifications, is required to model business processes in detail. The use of the Citizen's Charter does not suffice to model a business process. Having addressed how role dependencies in an organisation can improve service areas, further studies can extend to modelling of business processes. *i** framework is capable of capturing requirements in business processes. We plan to expand the organisational models to the modelling and evaluation of business processes to further improve service areas. We plan to apply frequency and time parameters within the service context in order to refine measurement in a finer context.

## Copyright